\documentclass{nature}
\usepackage{bbm}
\usepackage{amssymb}
\usepackage{amsmath}
\usepackage{amsfonts}
\usepackage{graphicx}
\usepackage{caption}
\usepackage{subfig}
\usepackage{txfonts}
\usepackage{epsfig}
\usepackage{color}
\usepackage{multirow}
\usepackage{mathrsfs}
 \usepackage{upgreek}
 \usepackage{bm}
 \usepackage{epstopdf}
 \usepackage[colorlinks=true,linkcolor=blue,citecolor=blue]{hyperref}%
\begin{document}

\title{Enhanced \bm{$\upalpha$} particle generation via proton-boron fusion reactions in laser-modulated plasma}
\author{Yihang Zhang$^{1}$, Zhe Zhang$^{1,4,5}$$^\star$, Yufeng Dong$^{1,4}$, Ke Fang$^{1}$, Haochen Gu$^{1,4}$, Yu Dai$^{1,4}$, Wei Qi$^{2}$, Zhigang Deng$^{2}$, Xiaohui Zhang$^{2}$, Lei Yang$^{2}$, Feng Lu$^{2}$, Zheng Huang$^{2}$, Kainan Zhou$^{2}$, Yuchi Wu$^{2}$, Weimin Zhou$^{2}$, Feng Liu$^{3,5}$, Guoqiang Zhang$^{6}$, Bingjun Li$^{7}$, Xu Zhao$^{3,5}$, Xiaohui Yuan$^{3,5}$, Chen Wang$^{8}$ and Yutong Li$^{1,4,5,9}$$^\star$}

\maketitle

\begin{affiliations}
\item
Beijing National Laboratory for Condensed Matter Physics, Institute of Physics, Chinese Academy of Sciences, Beijing 100190, China

$^{2}$ National Key Laboratory of Plasma Physics, Laser Fusion Research Center, CAEP, Mianyang, Sichuan 621900, China

$^{3}$ MoE Key Laboratory for Laser Plasmas and School of Physics, Shanghai Jiao Tong University, Shanghai 200240, China

$^{4}$ School of Physical Sciences, University of Chinese Academy of Sciences, Beijing 100049, China

$^{5}$ Collaborative Innovation Center of IFSA (CICIFSA), Shanghai Jiao Tong University, Shanghai 200240, China

$^{6}$ Shanghai Advanced Research Institute, Chinese Academy of Sciences, Shanghai 201210, China

$^{7}$ Department of Nuclear Science and Technology, Xi’an Jiaotong University, Xi’an 710049, China

$^{8}$ Shanghai Institute of Laser Plasma, China Academy of Engineering Physics, Shanghai 201800, China

$^{9}$ Songshan Lake Materials Laboratory, Dongguan, Guangdong 523808, China

$^\star$e-mail:zzhang@iphy.ac.cn; ytli@iphy.ac.cn

\end{affiliations}

\begin{abstract}
Aneutronic and nonradioactive properties make the proton-boron fusion a prospective candidate for fusion energy production through reactions following p+$^{11}$B$\rightarrow$3$\upalpha$ (p-$^{11}$B). 
However, it is difficult to achieve a thermal fusion ignition, since the low reaction cross-sections for center-of-mass energy below $\sim$100 keV. 
To realize fusion energy gain, it is essential to consider utilization of the maximum cross-section at the resonant peak of p-$^{11}$B fusion, and explore the nuclear reactions in plasma environment. 
In this work, p-$^{11}$B reactions triggered by interactions between energetic proton beams and laser-ablated boron plasma have been investigated. 
More than 200 times enhancement of $\upalpha$ particle emission efficiency (number ratio of escaping $\upalpha$ particles and boron nuclei) in plasma has been observed, compared with the cold boron. 
The proton beam transport path modulated by strong electro-magnetic fields in plasma could dominate the enhanced $\upalpha$ particle generation, due to a longer collisional length. 
In addition, an $\upalpha$ particle yield up to 1$\times$10$^{10}$ /sr has been measured via the pitcher-catcher scheme in plasma. %
This work could benefit understanding of the plasma effects on nuclear reaction dynamics, and also enable opportunities to explore physics in laser fusion associated with advanced fusion fuels.

\end{abstract}

Recent progress on magnetic\cite{Wurzel2022Progress, Baldzuhn2020Enhanced,Bozhenkov2020High} and inertial\cite{Abu2022Lawson,Jean2023First} confinement deuterium-tritium (DT) fusion has brought a major step forward in the relevant research area, and researchers has started to strive for higher energy gain in alternative advanced schemes, such as using aneutronic and nonradioactive fusion fuels. 
The proton-boron (p+$^{11}$B$\rightarrow$3$\upalpha$, p-$^{11}$B) reactions produces energetic $\upalpha$ particles without neutron generation.  
The abundant boron element in natural solid state makes target fabrication much easier, especially compared with the DT fusion which requires cryogenic technology and tritium breeding\cite{Reyes2016RecentDI}. 
However, the p-$^{11}$B reaction has much lower cross-section compared with the DT and deuterium-deuterium (DD) reactions in plasma of several or tens keV, so it is hard to realize self-sustained burning through thermal nuclear reactions. 
Fortunately, since the p-$^{11}$B reaction has cross-sections with intrinsic resonances predominant by the compound nucleus decay, it may have the potential to realize higher yields at the maximum resonant cross-section in avalanche processes\cite{Giuffrida2020High,Eliezer2016Avalanche}, via elastic nuclear collisions or intermediate reactions.

Using the technique of high power and intense laser for proton acceleration\cite{wagner2016maximum,Higginson2018Near} has opened a new realm for laser-driven nuclear physics\cite{Feng2022Femtosecond} and particle sources\cite{Roth2013Bright,Zhang2019Effects}. 
For $\upalpha$ particle generation from the p-$^{11}$B reactions, two main schemes have been studied: the “in target”\cite{becker1987low} and the “pitcher-catcher” schemes\cite{Bonvalet2021Energetic}. 
In the “in target” scheme, the longitudinal ponderomotive force of the intense laser pulse induces an electrostatic field that accelerates the protons from the target front surface to hundreds-keV. 
It was reported\cite{Giuffrida2020High} that up to 10$^{11}$ $\upalpha$ particles were achieved from a boron nitride target driven by a 600 J laser beam at an intensity of $\sim$10$^{16}$ W/cm$^{2}$. 
In “pitcher–catcher” scheme, the first target is used to generate energetic protons via the target normal sheath acceleration (TNSA), to enter and collide with the second target. 
The energetic protons have longer projectile range in the second target and more chances to collide with the boron nuclei. 
The record $\upalpha$ particle yield for the “pitcher–catcher” scheme\cite{Hegelich2023Photon} is up to 10$^{10}$ with a laser pulse of 80 J at 8×10$^{21}$  W/cm$^{2}$ . 
Because most of the $\upalpha$ particles are generated at a deep position inside the target, only a small fraction of the escaping ones could be directly detected and used for applications, due to the short stopping ranges of $\upalpha$ particles in a solid matter. 

Besides the recently impressive efforts made by the two principal schemes, back to 2013, a great achievement with laser produced p-$^{11}$B reactions has been made, and orders of magnitude enhancement for the fusion yields in plasma compared with a cold target has been presented\cite{Labaune2013Fusion}. 
Probable explanations of the production enhancement have been mentioned, as the pre-ionized atoms and pre-expelled electrons leading to less energy losses of the incident proton beams, as well as modified fusion cross-sections in plasma environments. 
Further convinced explanation has not been included in the work since the plasma effects on proton beam transportation and nuclear reaction dynamics\cite{Zhang2022Ion} is very complicated. 
However, these possible effects are essentially important for the ignition threshold assessment and optimum design for both magnetic and inertial confinement fusion, and are still investigated and debated so far.

As an aneutronic reaction, diagnostics for the only product $\upalpha$ particles from p-$^{11}$B fusion is also worth study. 
One major diagnostic is the CR-39 nuclear track detector \cite{nikezic2004formation}, but it is hard to distinguish different ion species and energy from the tracks with similar diameters. 
Another diagnostic is the Thomson parabola spectrometer\cite{thomson1911xxvi,carroll2010modified} with electric and magnetic dispersion for ions. 
But the tracks of $\upalpha$ particles could overlap with other ions possessing the same charge-to-mass ratio. 
The development of high-repetition-rate laser systems leads to multiple designs of on-line diagnostics, for example the ion time-of-flight (ToF) spectrometer\cite{Nakamura2006Real} based on scintillators and photomultiplier tubes. 
One of the main challenges of the ToF detector is to exclude signal from the strong $\upgamma$-rays driven by ultra-intense laser and solid interactions, by using robust electronic temporal gates\cite{Glebov2010National}. 
In addition, as we have explained in the “pitcher-catcher” scheme, diagnostics for $\upalpha$ particles stopped inside the targets can only rely on indirect ways, like measurements of photons from the parallel nuclear reactions\cite{Hegelich2023Photon}. 

In this paper, investigation of the p-$^{11}$B reactions in plasma has been reported, using the “pitcher-catcher” scheme with a fast proton beam and boron plasma triggered by two laser pulses. 
Three types of diagnostics have been employed to measure the $\upalpha$ particle production, which show consistency with one and another. 
The $\upalpha$ particle emission efficiency has been increased by more than 200 times (with about 20 times increasing on the total yields), compared with a cold boron target with an equivalent nucleus quantity.
It could be attributed to the proton beam transportation in the plasma affected by the spontaneous electro-magnetic fields. 

\section*{Results}
\subsection{Experimental setup.}

\begin{figure*}[!hbt]
\begin{center}
 \epsfig{file=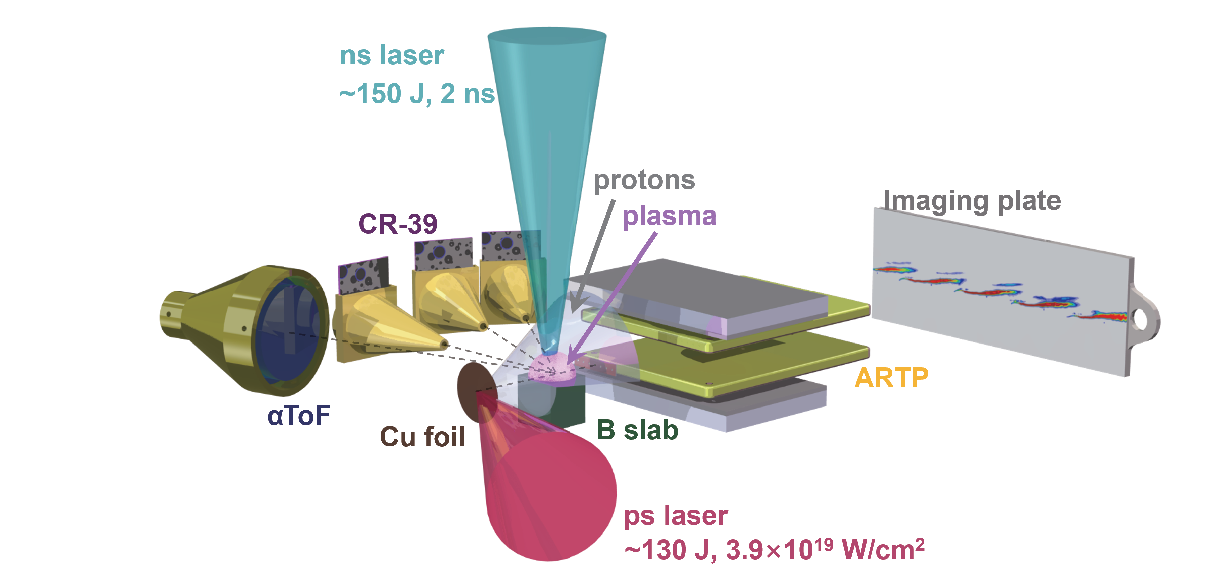,width=15.5cm}
 \end{center}
\caption{\textbf{The schematic diagram of the experimental setup.} The cubic boron slab and Cu foil were irradiated by the ns and ps laser beams, respectively, and the p-$^{11}$B reactions took place when the accelerated proton beam interacted with the boron plasma and un-heated boron symmetrically.
CR-39 and $\upalpha$-ToF detectors were set at different orientations to measure the $\upalpha$ particle yields and energy spectra.
The transmitted protons and the escaped $\upalpha$ particles from the plasma, entered the ARTP spectrometer through multi-pin-hole channels and their tracks were recorded by the BAS-TR Image Plate (IP) (see the colored map behind).}
\label{fig1}
\end{figure*}

The experiment on p-$^{11}$B nuclear reactions in plasma was carried out at the XG-III laser facility\cite{zhu2017xingguang}. 
As shown in Fig.~\ref{fig1}, a solid boron target as a 2-mm cube was irradiated by a nanosecond (ns) laser, delivering 150 J to the top surface in a 2-ns square pulse. 
Then a 10-$\upmu$m-thick Cu foil was irradiated by a picosecond (ps) laser with a 20-$\upmu$m FWHM spot at the best focus and 0.75 ps pulse length, giving an intensity of 3.9×10$^{19}$ W/cm$^{2}$. 
The energetic proton beam was driven dominantly by the TNSA, then traversed the boron target, inducing p-$^{11}$B nuclear reactions via the “pitcher-catcher” mechanism. 
The central axis of the irradiated boron surface was at the same direction with the normal of the Cu foil. 
Thus, the proton beam has symmetrically bisecting distribution about the boron surface, interacting with the boron solid and plasma simultaneously. 
To demonstrate the reactions in different plasma conditions, the time delay between the ns and ps lasers has been changed from 0.3 ns to 2.1 ns in the experiment, with a jitter within 0.1 ns.

Since the $\upalpha$ particles are often emitted accompanied by other ions, here, in order to measure their yields, we have employed three types of detectors to recognize different ion species. 
Firstly, three pieces of CR-39 detectors were set in the horizontal plane, at angles of 70°, 80°, and 90°, respectively, from the normal of the Cu foil. 
Secondly, one set of $\upalpha$ time-of-flight detectors ($\upalpha$-ToF) at 100.9° from the target normal, was implemented to provide the energy spectra. 
Thirdly an angular resolved Thomson parabola (ARTP) spectrometer\cite{zhang2018angular} was employed to diagnose energy spectra of forward $\upalpha$ particles, with a 13-$\upmu$m-thick Al filter for ions with the same mass-to-charge ratio. 
Besides, the ARTP spectrometer also measured the protons after penetrating the boron plasma through different angle of path within -8°$\sim$8°, carrying information of interactions with the plasma. 
The proton beam pattern was recorded by a stack of Radiochromic films (RCFs)\cite{vatnitsky1997radiochromic} (not shown in Fig.~\ref{fig1} for brevity) in front of the ARTPS configured for plasma imaging. 

\subsection{Experimental results.}
\begin{figure*}[!hbt]
\begin{center}
 \epsfig{file=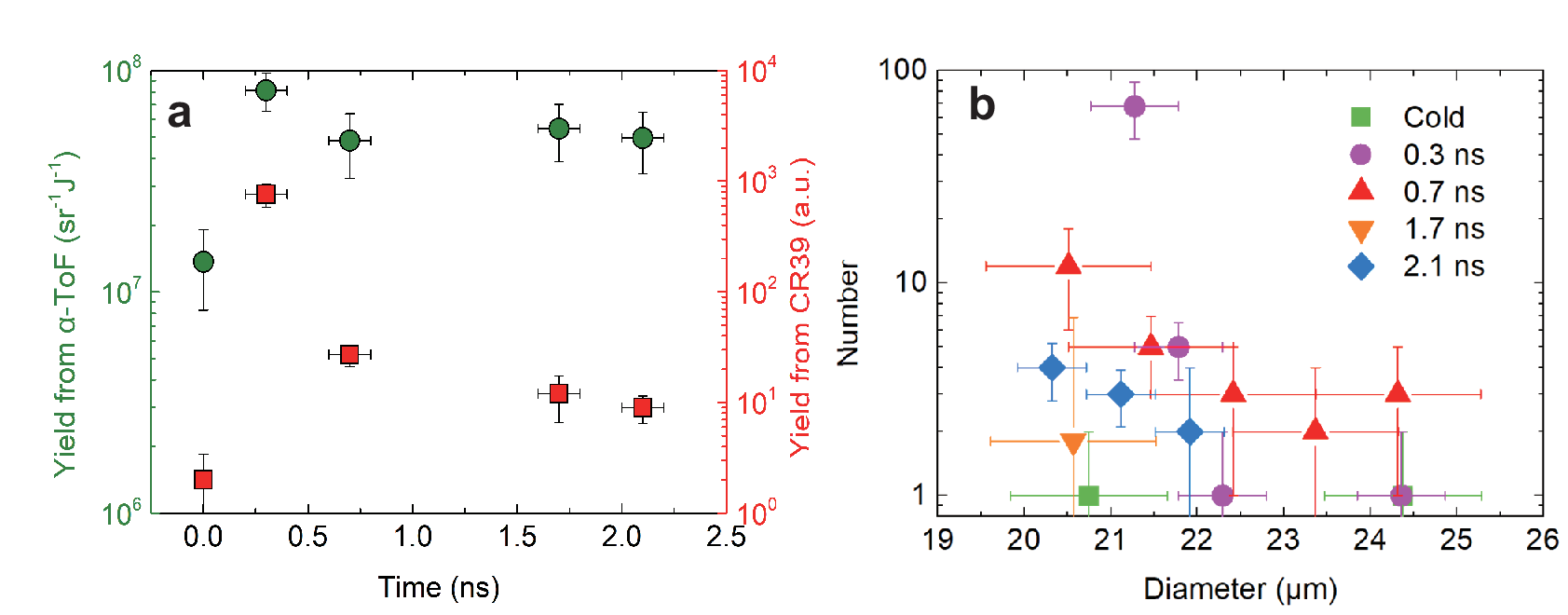,width=15cm}
 \end{center}
\caption{\textbf{Emitted $\upalpha$ particle yields.} \textbf{a} Variations of the $\upalpha$ particle yields as the boron target evolves from a cold target to plasma, derived from the $\upalpha$-ToF (green) and CR-39 detectors (red), normalized by the collecting solid angles and the ps laser energy. 
\textbf{b} Numbers of $\upalpha$ particle tracks on the CR-39 detectors in the penumbral edge with diameters from 20 to 25 $\upmu$m, at different boron pre-heated time. 
The error bars in the longitudinal direction are from the systematic errors of the three CR-39 detectors.}
\label{fig2}
\end{figure*}

From the $\upalpha$-ToF detector, the $\upalpha$ particle yields normalized to the ps laser energy have variation as the boron evolving from a cold target to plasma, which is shown as the green circles in Fig.~\ref{fig2}a. 
The time is defined as the delay from the rising edge of the ns laser to the ps pulse. 
The total numbers of the recognized $\upalpha$ particles on the three CR-39 detectors are plotted in Fig.~\ref{fig2}a as the red squares, which shows a qualitatively similar but more tremendous variation compared with the results from the $\upalpha$-ToF. 
This is owing to the signal lost on the saturation area of the CR-39 due to numerous exhaust protons (see Methods). 
The CR-39 results in Fig.~\ref{fig2} are recognized around the penumbral edge of the pin-hole imaging pattern, which have positive proportion with the total particle numbers.
Figure~\ref{fig2}b shows the $\upalpha$ particle numbers with different track diameters recorded around the CR-39 penumbral edge with different laser and target conditions. 
According to the calibration and ion stopping calculation\cite{S2003Spectrometry}, the charactered diameter of the $\upalpha$ particle track is less than 25 $\upmu$m under our etching condition (see Methods), which could be partly distinguished from protons with diameters less than 20 $\upmu$m. 
For the experimental shots only with the ns laser ablating the boron target (without proton beams), neither of the two diagnostics see any signal from $\upalpha$ particles, which indicated the thermal nuclear reaction driven by the ns laser can be negligible. 
According to the calibrated $\upalpha$-ToF detector, from the interactions between proton beams and cold boron targets, the $\upalpha$ particle yield is 1.4×10$^7\pm$4.4×10$^6$ /sr/J, with error bars summing systematic and statistical errors in quadrature. 
At $t$ = $t_0$ + 0.3 ns (where $t_0$ is the arrival time of ns laser pulse at the boron target) the yield has been increased to 8.1×10$^7\pm$2.8×10$^6$ /sr/J, which is at the similar level and slightly higher than the record yield of laser driven p-$^{11}$B reaction through the pitcher-catcher scheme ($\sim$5×10$^7$ /sr/J with the $\sim$2×10$^{21}$ W/cm$^2$ Texas Petawatt laser pulse\cite{Hegelich2023Photon}). 

\begin{figure*}[!hbt]
\begin{center}
 \epsfig{file=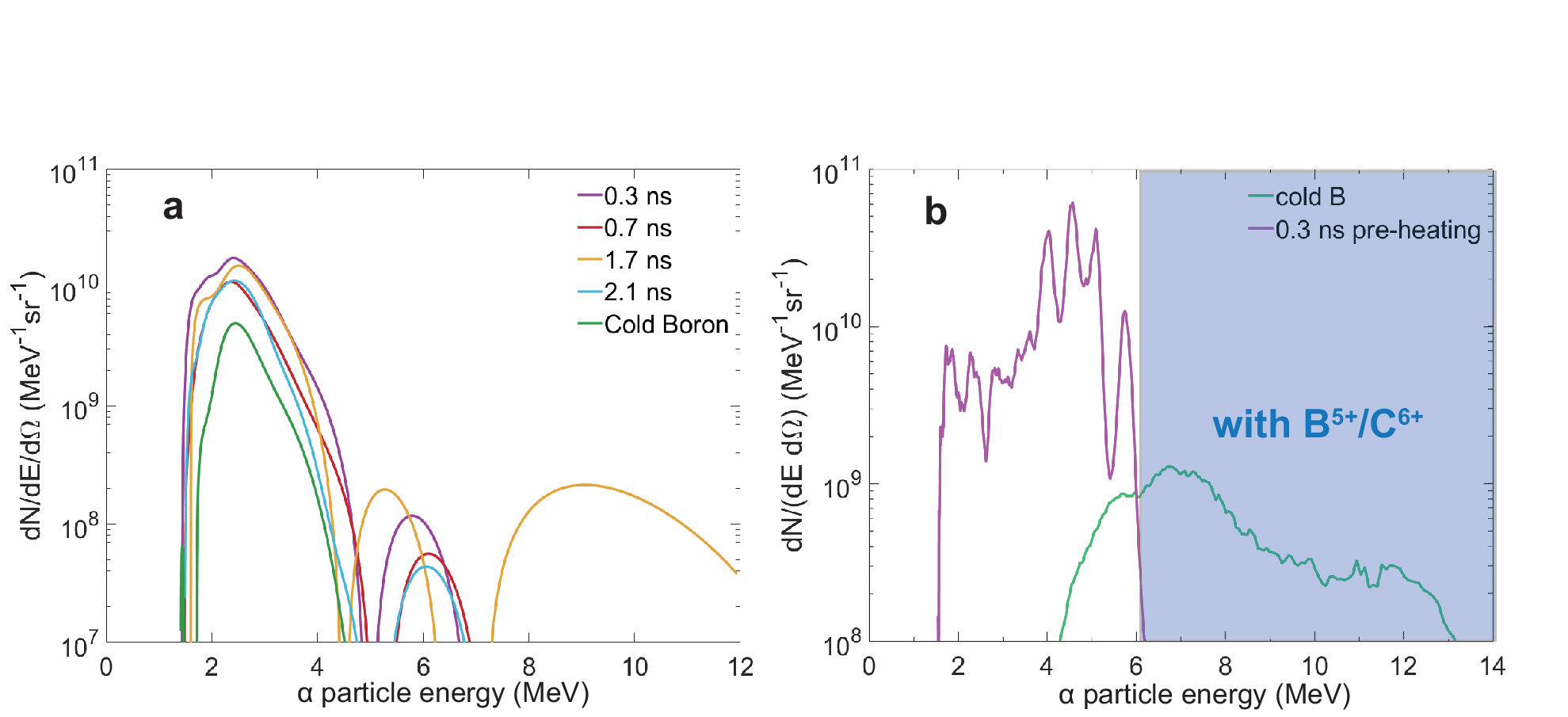,width=15cm}
 \end{center}
\caption{\textbf{$\upalpha$ particle energy spectra.} \textbf{a} The $\upalpha$ particle energy spectra measured by the $\upalpha$-ToF detector with different boron pre-heating time, as well as the cold boron target. \textbf{b} $\upalpha$ particle energy spectra comparing the cases of 0.3 ns pre-heating and the cold boron target, measured by the ARTP spectrometer. 
In the ARTP spectrometer, the $\upalpha$ particles with energy higher than 6 MeV induce signal overlapped with other boron and carbon ions. 
}
\label{fig3}
\end{figure*}

The statistic of track diameters on the CR-39 presented in Fig.~\ref{fig2}b indicates the energy distribution of the $\upalpha$ particles varies with the plasma conditions. 
The energy spectra can be solved from the $\upalpha$-ToF detector, according to the flight time of $\upalpha$ particles with different velocities, which are plotted in Fig.~\ref{fig3}a. 
It shows the fusion production in different boron plasma with the proton beam driven at different time. 
In the meantime, the green curve has set a reference for the $\upalpha$ spectrum from protons interacted with a cold boron target. 
There is increased production from the cold target evolving to plasma. 
Meanwhile, in the plasma conditions, $\upalpha$ particles with energy higher than 5 MeV has been detected by the $\upalpha$-ToF detector. 
However, in the solid target, these high energy $\upalpha$ particles have less number and higher energy loss. 
The highest yield occurs with the shortest time delay of 0.3 ns, and there is an enhancement by factor of $\sim$5.9 compared with the cold target. 
Spectra of $\upalpha$ particles close to the target normal direction could be derived from the results of the ARTP spectrometer, as shown in Fig.~\ref{fig3}b. 
In the case of the cold boron target, the ARTP spectrometer could hardly observe $\upalpha$ particles due to the stopping of the 2-mm thick boron bulk, and the signal of forward $\upalpha$ particles with small angle has just exceeded the background noise.

\section*{Discussions}
The projectile range of an $\upalpha$ particle with 2.9 MeV (the averaged kinetic energy of the three $\upalpha$ particles from a p-$^{11}$B reaction) in a solid boron target is about 8.1 $\upmu$m, which is the same with a 0.755 MeV proton beam. 
For protons with higher energy and inducing fusion at deeper positions, the $\upalpha$ particle could hardly escaped from the target through the backward direction. 
We have demonstrated fusion reactions generated from a proton beam and solid boron target with Monte-Carlo simulations using the Geant4 code\cite{Agostinelli2003G}. 
The simulation has adopted the experimentally measured energy and divergence distributions of the proton beams, and the same geometry of the beam-target interactions and detector position with the experiment (see Methods). 
The simulation shows the total $\upalpha$ particle number is 10 times of the escaping ones from the target. 
As a solid material evolves to plasma state, the stopping range of fast ions would be changed and might result in different escaping $\upalpha$ particle numbers. 
However, the ns laser ablation area is at the center of the 2 mm$\times$2 mm boron top surface, far from the proton projectile range. 
So the differences of the measured $\upalpha$ particle numbers, between the cases with and without the ns ablation, may come from the contribution in the boron plasma, with very few additional escaping particles from the solid target.

\begin{figure*}[!hbt]
\begin{center}
 \epsfig{file=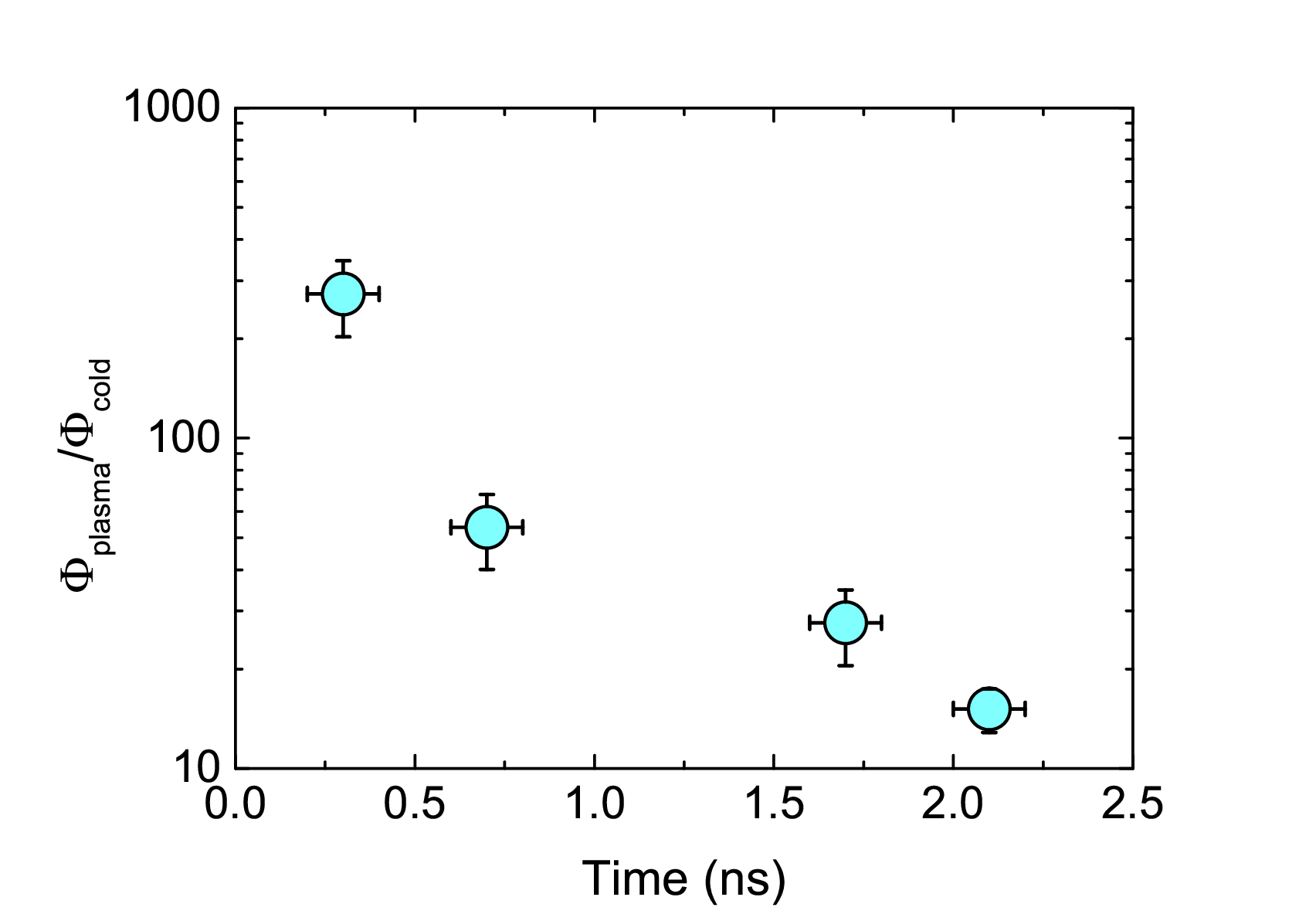,width=10cm}
 \end{center}
\caption{\textbf{The $\upalpha$ particle emission efficiency in plasma compared with the cold target.} Here the $\upalpha$ particle number haves been normalized to the ps laser energy. 
The $\upalpha$ particle numbers generated in plasma are deduced from the total detected numbers subtracted by the one detected in the cold target situation. 
The boron nucleus numbers in plasma has been inferred from the 3D Flash simulations.}
\label{fig4}
\end{figure*}

The $\upalpha$ particle emission efficiency (fraction of burnt and escaped nuclei) can be defined as the ratio $\Phi$ = $N_{esc~\alpha}/N_B$ between the number of escaping $\upalpha$ particles $N_{esc~\alpha}$ and the one of boron nuclei $N_B$ interacted with the proton beam. 
According to the geometry of the experimental setup, it is assumed that the protons interact with the blow-off boron plasma and the solid boron slab symmetrically. 
The boron plasma ablated by the laser pulse has been calculated using the hydrodynamic simulations with the three-dimensional (3D) FLASH code\cite{Fryxell2000FLASH}. 
Figure~\ref{fig4} shows the $\upalpha$ particle emission efficiency $\Phi$ in plasma divided by the one in the cold boron, which evolves with the time delay between the ablated boron and the proton beam. 
The effective cold boron number is deduced from reaction cross-section weighted volume of the solid target penetrated by protons, according to the Geant4 simulations, and the corresponding mass is 20 $\upmu$g.

The p-$^{11}$B reaction numbers d$N_{fus}$ generated with per unit number of incident proton d$N_{p}$ along its propagation distance $l$ in a target with boron nucleus density of $n_B$ can be written as
\begin{eqnarray}
\frac{\mathrm{d}N_{fus}}{\mathrm{d}N_p}=n_B\int^l_0\sigma \mathrm{d}r
\end{eqnarray}
where $\sigma$ is the reaction cross-section.
To our best knowledge the cross-section barely changes in this experimental laser and plasma conditions\cite{Salpeter1954ElectronSA,Lv2022Enhanced}. 
The blow-off plasma has much lower areal density than the cold target, and most of the $\upalpha$ particles can escape from the plasma. 
Even if the $\upalpha$ particles generated in the solid boron could totally escaped, it could not account for the hundreds times enhancement of emission efficiency shown in Fig.~\ref{fig4}. 
One of the probable reasons for the enhancement on the $\upalpha$ particle yields is the lengthening of proton path length $l$ (i.e., the transmitting time) inside the plasma. 

\begin{figure*}[!hbt]
\begin{center}
 \epsfig{file=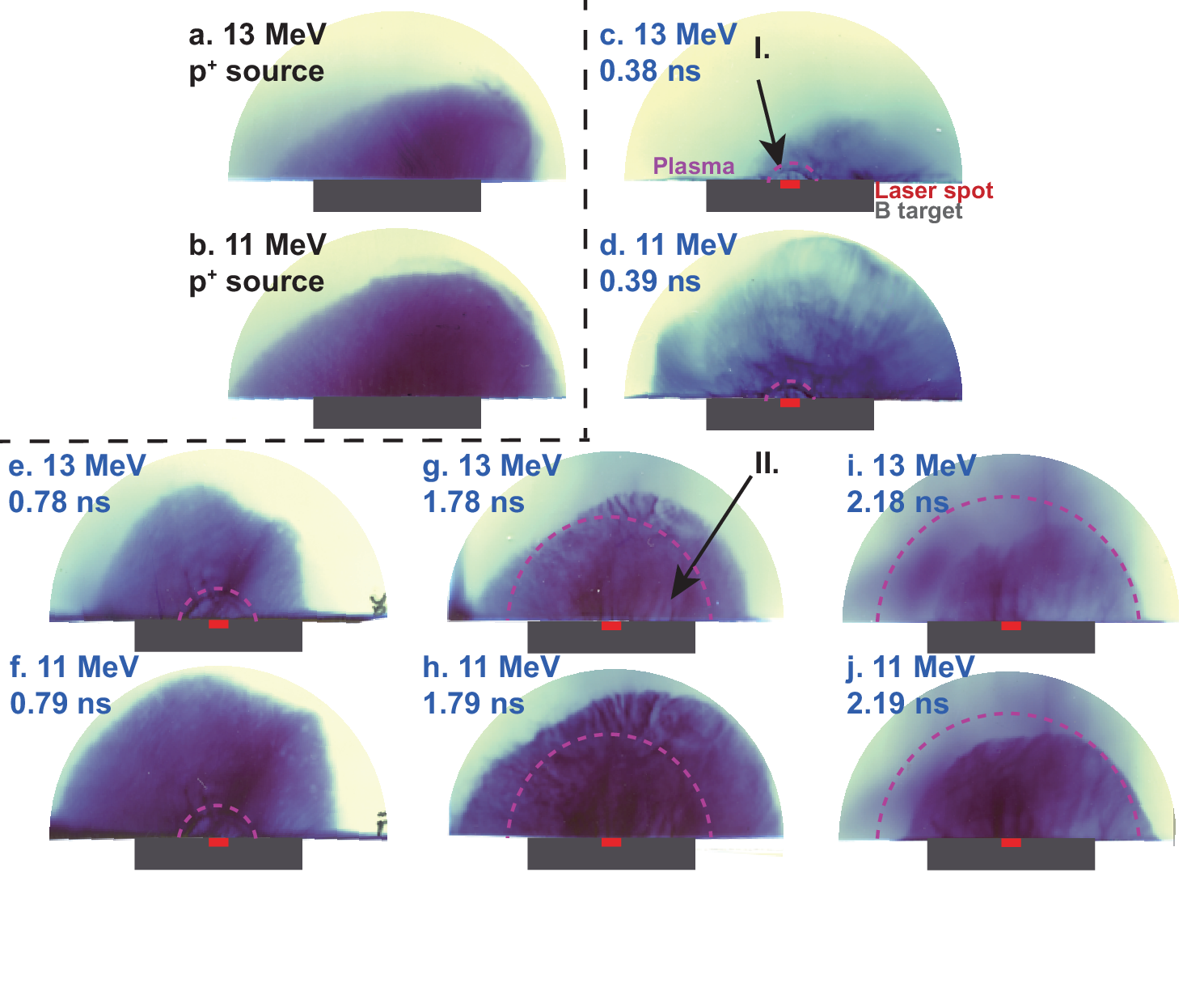,width=14cm}
 \end{center}
\caption{\textbf{Images on the RCF stacks for 11 and 13 MeV protons.} 
The angular distribution of the beam source are shown in \textbf{a} and \textbf{b}, and the radiographs at different ns-ps time delay are in \textbf{c-j}. 
The darker regions correspond to stronger proton fluxes. 
The grey and red symbols represent the initial boron target and the laser ablation positions. 
The magenta dashed lines are the plasma expansion boundary according to the FLASH simulations.
}
\label{fig5}
\end{figure*}

To diagnose the proton beam transmission inside the blow-off boron plasma, side-on proton-probed imaging has been set, using the RCF stacks. 
A half beam spot of the proton source has been characterized without the ns laser ablation, presented in Fig.~\ref{fig5}a and b for the 13 MeV and 11 MeV protons, respectively. 
It shows the proton angular distribution before interactions with the boron target could be regular and smooth. 
The beam patterns after the interactions with the plasma at different time are shown in the other panels of Fig.~\ref{fig5}. 
Here the grey and red squares indicate the initial position of the boron target and the ablating laser spot, while the magenta dashed lines are the blow-off plasma boundary deduced from the FLASH simulations. 
At around 0.4 and 0.8 ns (Fig.~\ref{fig5}c-f), bubble-like structures associated with the blow-off plasma can be seen, noted as Region I. 
These regions for modulation of probe proton density reveal Biermann magnetic field induced by unparallel gradients of electron density $n_e$ and temperature $T_e$, or strong electric field driven by plasma pressure $P_e$ gradients during unstable ablation.
At later time as around 1.8 ns and 2.2 ns (Fig.~\ref{fig5}g-j), many radial filament structures have also been observed in Region II., which could come from the complex electromagnetic field striations. 
These fields may be generated by thermal driven Weibel-like instability caused by anisotropic and localized heat-absorption\cite{Estabrook1978Qualitative,Willi1981Optical}, or generated according to the Biermann-Battery effect in the growing hydrodynamic instabilities\cite{Gao2012Magnetic}. 
In addition, considering the unsmoothed ns laser in our experiment could have modulated structures in the beam spot, filamentation and perturbation of the plasma density may grow. 
The laser filaments of lower density may be surrounded by a density gradient across the filament and a temperature gradient along it, and $\nabla n_e\times\nabla T_e$ gives rise to the magnetic fields. 

\begin{figure*}[!hbt]
\begin{center}
 \epsfig{file=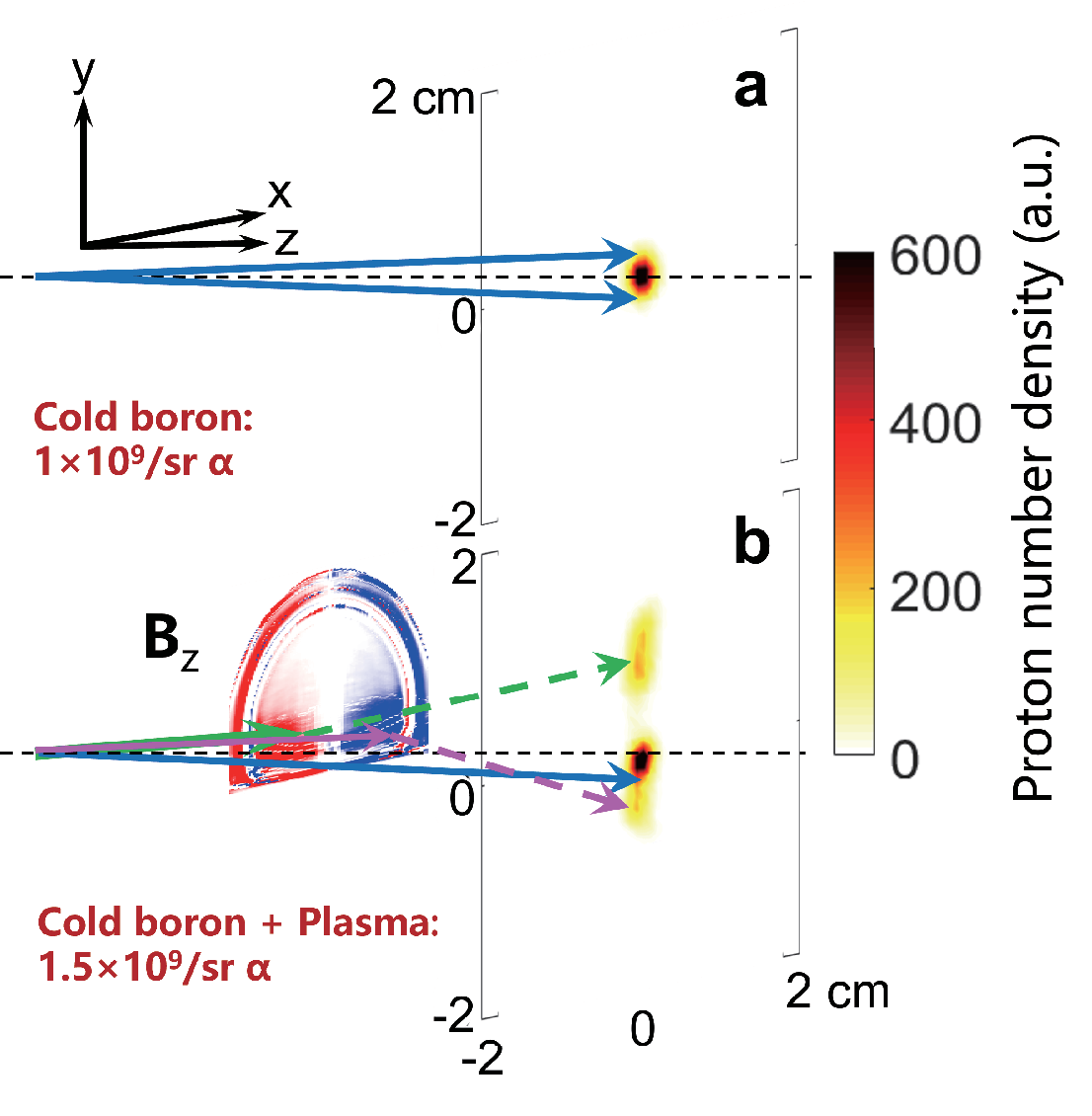,width=10cm}
 \end{center}
\caption{\textbf{Proton beam spots on the detecting plane simulated by the Geant4 code.} 
The proton beam has energy of 4 MeV a divergence angle of 9°. \textbf{a} The protons transport in the gas without magnetic fields. 
\textbf{b} With a toroidal magnetic field of 350 T at maximum.
The solid arrows indicate the proton directions, while the dashed ones are showing the displacements inducing by the magnetic fields.
(The distances are shown as a principle scheme and not in the real scale.)
}
\label{fig6}
\end{figure*}

For qualitatively understanding the proton trajectory in the self-generated fields, using the Geant4 code, the detected proton beam patterns after the magnetic field induced deflection have been simulated. 
In the simulation the proton beam propagates through a low density boron gas, and hits a detecting plane. 
The gas is in a volume of 2-mm cubic with uniform density of 0.02 g/cm$^3$. 
We take the toroidal magnetic field driven by $\nabla n_e\times\nabla T_e$ during the plasma expansion as an example. 
The 3D distributed magnetic field from the Flash simulation (at 0.8 ns when the plasma has expanded for about 800 $\upmu$m away from the initial surface) is post-processed in the Geant4 code. 
The distance between the proton source and the center of the boron gas is 4 mm, while the detector plane is 60 mm away from the gas, which are the same with the experimental conditions. 
The results are presented in Fig.~\ref{fig6}, where the boron gas (not shown) is located on top of the $x$-$z$ plane. 
The detected proton beam spot without the field is shown in Fig.~\ref{fig6}a. 
When transversely interacting with a toroidal magnetic field, as presented in Fig.~\ref{fig6}b, the protons can be deflected away from or close towards the solid boron slab, depending on the specific field strength they experience, which may leads to longer interaction lengths or higher target density, and could give rise to the fusion reaction yield. 
After the cold boron solid target has been added beneath the $x$-$z$ plane, the simulation shows an 1.5 times enhancement on the escaping $\upalpha$ particle numbers with the magnetic field in the gas, compared to the situation without the field.
It should be noted that the peak magnetic field strength is only about 4 T in the hydrodynamic simulations, but in the Monte-Carlo simulation it has been amplified artificially to 350 T at maximum. 
Moreover, in the real ablated plasma, there is much more intricate field distribution indicated from Fig.~\ref{fig5} involving the similar physical processes of proton trace wiggling, which may cause more obvious change on the reaction yield.

In summary, p-$^{11}$B reactions initiated by laser-accelerated proton beams in laser-produced boron plasma has been studied. 
Our experimental results show more than 200 times enhanced $\upalpha$ particle emission efficiency in boron plasma, compared with a cold boron target. 
The electro-magnetic fields generated by the laser ablation and filamentation instability may alter the proton traces in the plasma and enhance the collision probability between protons and boron nuclei. 
In addition, the under-dense plasma provides lower stopping power for the $\upalpha$ particles, and can further increase the escaping particles numbers. 
In future research, proper choosing for the proton energy spectrum and target areal density (with advanced materials such as low-density foams or nano-wires) could be adopted to optimize the $\upalpha$ particle yield. 
This work could benefit understanding of the plasma effects on $\upalpha$ particle generation in the p-$^{11}$B fusion reactions, which plays a key role in the advanced fusion reactors.

\section*{Methods}
\subsection{CR-39 detectors.} For fusion yield diagnosing, three pieces of CR-39 detectors were set to record the tracks of $\upalpha$ particles, using conical collimators with pin-holes to shield it from scattered ions. 
The collimators were made by 0.1-mm-thick Al conical shells, with 0.5-mm-diameter pinhole tips to collect particles emitted from the target. 
There was a cover of 5-$\upmu$m-thick Al filter on the CR-39 to block the $\upalpha$ particles below 1.6 MeV and protons below 0.5 MeV. 
The particle tracks on the CR-39 have been read after 6-h etching by 6.25 molarity NaOH at 80°C. 
In the intense laser-plasma interactions induced by the ps laser, numerous protons exhaust due to charge separation, which made the particle tracks heavily overlapped at the central area of the CR-39. 
So the data was counted according to the unsaturated tracks on the edge.
\subsection{$\upalpha$-ToF detectors}. The $\upalpha$ Time-of-Flight ($\upalpha$-ToF) detectors are based on plastic scintillators and microchannel plate photomultiplier tubes (MCP-PMT)\cite{Glebov2010National}. 
Here a 1-mm-thick EJ212 plastic scintillator was used, and the $\upalpha$ particles with energy less than 10 MeV could be stopped in the scintillator and the energy deposition transferred to photoelectrons. 
The MCP was gated off to prevent the PMT from saturation and overdrive from the strong $\upgamma$-rays and its afterglow on the scintillators. 
The $\upalpha$ particle numbers was deduced from quantitative calibration of the detector using the $\upalpha$ decay of $^{241}$Am.

\subsection{ARTP spectrometer}. The design and principle of the angular resolved Thomson parabola (ARTP) spectrometer have been reported in Ref. 24.
The central incident pin-hole of the spectrometer was set at the normal of the Cu foil target, in order to measure the proton from TNSA. 
The spectrometer also caught $\upalpha$ particles, since the distance from the target is only 78 mm with a collection solid angle of 7.6×10$^{-7}$ sr. 
And a 13-$\upmu$m-thick Al foil was used as a filter for heavy ions. 
According to Monte-Carlo simulations, for ions with mass-to-charge ratio A/Z = 2, the B$^{5+}$ and C$^{6+}$ ions with energy higher than 20 MeV and 25 MeV could be recorded by the imaging plate in the ARTP spectrometer, but the tracks have no overlap with the $\upalpha$ particles.
\subsection{Plasma characterization.} 
\begin{figure*}[!hbt]
\begin{center}
 \epsfig{file=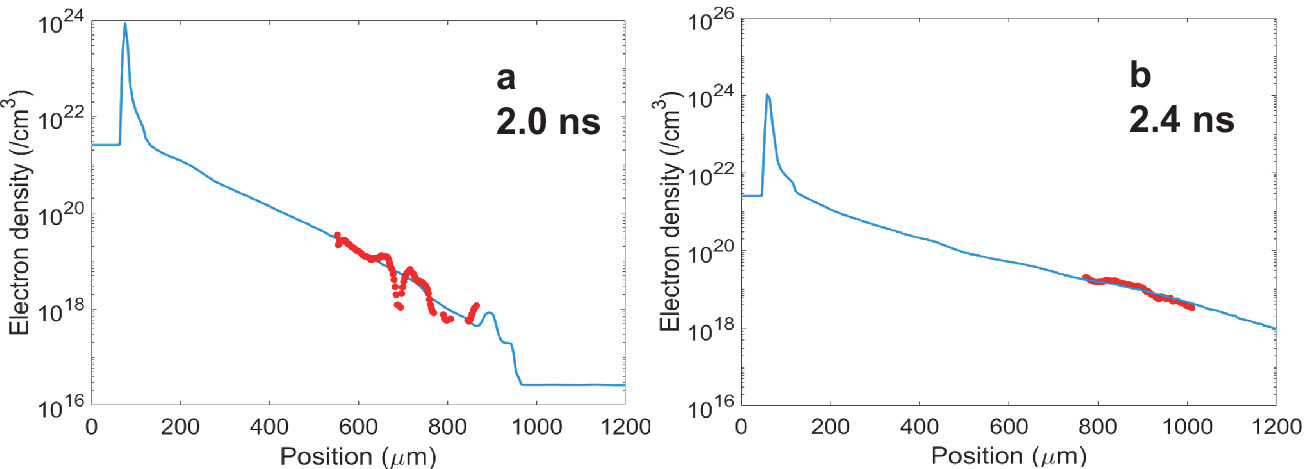,width=15cm}
 \end{center}
\caption{\textbf{Longitudinal electron density distribution along the axis of target normal.} 
Comparison between measured (red circles) and simulated (blue curves) of free electron density along the axis of target normal, at 2.0 ns ($\textbf{a}$) and 2.4 ns ($\textbf{b}$) after the rising edge of the ns laser.
}
\label{fig7}
\end{figure*}

A 400 nm probe laser beam with an energy of 3 mJ and duration of 50 fs has propagated parallelly to the boron target surface which irradiated by the nanosecond (ns) laser, to measure the plasma density distribution. 
2D phase maps of the probe beam through the plasma were recorded by the SID4-HR sensor\cite{SID4-HR}, and the plasma density can be inverted, with a timing scan in different shots. 
The FLASH code has been used to simulate the boron plasma evolution driven by the ns laser. 
Typical longitudinal distribution of plasma density at 2.0 ns and 2.4 ns after the rising edge of the ns pulse in Fig.~\ref{fig7} shows nice agreement between the measurement and simulation in the detectable density range.

After this cross-calibration, we use the FLASH simulation results to calculate the boron nucleus number $N_B$ ablated from the solid target. 
And then the $\upalpha$ particle emission efficiency $Phi$ can be deduced.

\subsection{Fusion reaction simulations.} 
\begin{figure*}[!hbt]
\begin{center}
 \epsfig{file=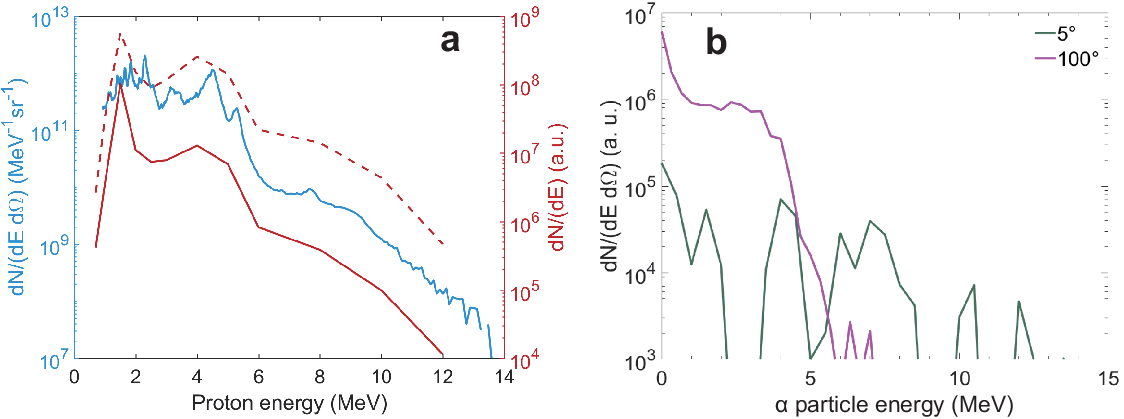,width=15cm}
 \end{center}
\caption{\textbf{Monte-Carlo simulation results of $\upalpha$ particles generated in the cold boron target.} 
\textbf{a} Forward proton energy spectrum measured by the central pin-hole of the ARTP spectrometer (blue curve) and the total (red dash curve) and escaped (red solid curve) $\upalpha$ particle numbers from the Geant4 simulations. \textbf{b} The $\upalpha$ particle energy spectra in 5° and 100° from the Geant4 simulations.
}
\label{fig8}
\end{figure*}

The Geant4 code with fusion cross section data TENDL2021\cite{tendl} has been used to simulate $\upalpha$ particle generation from the p-$^{11}$B fusion. 
The same geometry with the experimentally used the proton beam and the boron solid has been taken in the simulation. 
The ARTP spectrometer has provided fine proton energy spectra in an angular range of -8°$\sim$8°, while the RCF stack has information of the whole pattern of the proton beam with relatively coarse resolution. 
The simulation uses the results of the proton beam according to both of the diagnostics. 
The energy spectrum of the forward proton beam at 0 degree is shown in the blue curve of Fig.~\ref{fig8}a. 
The $\upalpha$ particle numbers generated with different proton energy are shown in the red curves, where the solid and dashed ones indicate escaped and total number of $\upalpha$ particles, respectively. 
The highest reactivity comes from the moderate energetic protons due to the reaction probability and the proton energy spectra. 
Figure~\ref{fig8}b shows the escaped $\upalpha$ particle energy spectra from the Geant4 code in 5° and 100° with respect to the proton beam direction, which simulate the measurements by ARTP spectrometer and $\upalpha$-ToF, respectively. 
The energy spectra are consistent with the experimental results for the cold target in Fig.~\ref{fig3}, except for particles with very low energy filtered by the diagnostics.

\begin{addendum}
\item [Acknowledgement]
This study was supported in part by the National Natural Science Foundation of China (Grant Nos. 11827807 and 12105359) and the Strategic Priority Research Program of the Chinese Academy of Sciences (Grant Nos. XDA25010100, XDA25010300, and XDA25030100).


\item [Competing Interests]
The authors declare that they have no competing financial interests.

\item [Correspondence]
Correspondence and requests for materials should be addressed to Zhe Zhang and Yutong Li.

\end{addendum}

\nocite{*}
\bibliographystyle{naturemag} 
\bibliography{sample}

\begin{thebibliography}{10}
\expandafter\ifx\csname url\endcsname\relax
  \def\url#1{\texttt{#1}}\fi
\expandafter\ifx\csname urlprefix\endcsname\relax\def\urlprefix{URL }\fi
\providecommand{\bibinfo}[2]{#2}
\providecommand{\eprint}[2][]{\url{#2}}

\bibitem{Wurzel2022Progress}
\bibinfo{author}{Wurzel, S.~E.} \& \bibinfo{author}{Hsu, S.~C.}
\newblock \bibinfo{title}{Progress toward fusion energy breakeven and gain as
  measured against the lawson criterion}.
\newblock \emph{\bibinfo{journal}{Phys. Plasmas}}
  \textbf{\bibinfo{volume}{29}}, \bibinfo{pages}{062103}
  (\bibinfo{year}{2022}).

\bibitem{Baldzuhn2020Enhanced}
\bibinfo{author}{Baldzuhn, J.} \emph{et~al.}
\newblock \bibinfo{title}{Enhanced energy confinement after series of pellets
  in wendelstein 7-x}.
\newblock \emph{\bibinfo{journal}{Plasma Phys. Control. Fusion}}
  \textbf{\bibinfo{volume}{62}}, \bibinfo{pages}{055012}
  (\bibinfo{year}{2020}).

\bibitem{Bozhenkov2020High}
\bibinfo{author}{Bozhenkov, S.} \emph{et~al.}
\newblock \bibinfo{title}{High-performance plasmas after pellet injections in
  wendelstein 7-x}.
\newblock \emph{\bibinfo{journal}{Nucl. Fusion}} \textbf{\bibinfo{volume}{60}},
  \bibinfo{pages}{066011} (\bibinfo{year}{2020}).

\bibitem{Abu2022Lawson}
\bibinfo{author}{Abu-Shawareb, H.} \emph{et~al.}
\newblock \bibinfo{title}{Lawson criterion for ignition exceeded in an inertial
  fusion experiment}.
\newblock \emph{\bibinfo{journal}{Phys. Rev. Lett.}}
  \textbf{\bibinfo{volume}{129}}, \bibinfo{pages}{075001}
  (\bibinfo{year}{2022}).

\bibitem{Jean2023First}
\bibinfo{author}{Nicola, J.-M. G.~D.}
\newblock \bibinfo{title}{{First demonstration of fusion ignition by inertial
  confinement fusion (ICF) at the National Ignition Facility (NIF) at LLNL}}.
\newblock In \emph{\bibinfo{booktitle}{High Power Lasers for Fusion Research
  VII}}, vol. \bibinfo{volume}{PC12401}, \bibinfo{pages}{PC1240101}.
  \bibinfo{organization}{International Society for Optics and Photonics}
  (\bibinfo{publisher}{SPIE}, \bibinfo{year}{2023}).

\bibitem{Reyes2016RecentDI}
\bibinfo{author}{Reyes, S.} \emph{et~al.}
\newblock \bibinfo{title}{Recent developments in ife safety and tritium
  rerecentsearch and considerations for future nucl. fusion facilities}.
\newblock \emph{\bibinfo{journal}{Fusion Eng. Des.}}
  \textbf{\bibinfo{volume}{109}}, \bibinfo{pages}{175--181}
  (\bibinfo{year}{2016}).

\bibitem{Giuffrida2020High}
\bibinfo{author}{Giuffrida, L.} \emph{et~al.}
\newblock \bibinfo{title}{High-current stream of energetic \ensuremath{\alpha}
  particles from laser-driven proton-boron fusion}.
\newblock \emph{\bibinfo{journal}{Phys. Rev. E}}
  \textbf{\bibinfo{volume}{101}}, \bibinfo{pages}{013204}
  (\bibinfo{year}{2020}).

\bibitem{Eliezer2016Avalanche}
\bibinfo{author}{Eliezer, S.}, \bibinfo{author}{Hora, H.},
  \bibinfo{author}{Korn, G.}, \bibinfo{author}{Nissim, N.} \&
  \bibinfo{author}{Martinez~Val, J.~M.}
\newblock \bibinfo{title}{{Avalanche proton-boron fusion based on elastic
  nuclear collisions}}.
\newblock \emph{\bibinfo{journal}{Phys. Plasmas}} \textbf{\bibinfo{volume}{23}}
  (\bibinfo{year}{2016}).

\bibitem{wagner2016maximum}
\bibinfo{author}{Wagner, F.} \emph{et~al.}
\newblock \emph{\bibinfo{journal}{Phys. Rev. Lett.}}
  \textbf{\bibinfo{volume}{116}}, \bibinfo{pages}{205002}
  (\bibinfo{year}{2016}).

\bibitem{Higginson2018Near}
\bibinfo{author}{Higginson, A.} \emph{et~al.}
\newblock \bibinfo{title}{Near-100 mev protons via a laser-driven
  transparency-enhanced hybrid acceleration scheme}.
\newblock \emph{\bibinfo{journal}{Nat. Commun.}} \textbf{\bibinfo{volume}{9}},
  \bibinfo{pages}{724}.

\bibitem{Feng2022Femtosecond}
\bibinfo{author}{Feng, J.} \emph{et~al.}
\newblock \bibinfo{title}{Femtosecond pumping of nuclear isomeric states by the
  coulomb collision of ions with quivering electrons}.
\newblock \emph{\bibinfo{journal}{Phys. Rev. Lett.}}
  \textbf{\bibinfo{volume}{128}}, \bibinfo{pages}{052501}
  (\bibinfo{year}{2022}).

\bibitem{Roth2013Bright}
\bibinfo{author}{Roth, M.} \emph{et~al.}
\newblock \bibinfo{title}{Bright laser-driven neutron source based on the
  relativistic transparency of solids}.
\newblock \emph{\bibinfo{journal}{Phys. Rev. Lett.}}
  \textbf{\bibinfo{volume}{110}}, \bibinfo{pages}{044802}
  (\bibinfo{year}{2013}).

\bibitem{Zhang2019Effects}
\bibinfo{author}{Zhang, Y.} \emph{et~al.}
\newblock \bibinfo{title}{Effects of internal target structures on laser-driven
  neutron production}.
\newblock \emph{\bibinfo{journal}{Nucl. Fusion}} \textbf{\bibinfo{volume}{59}},
  \bibinfo{pages}{076032} (\bibinfo{year}{2019}).

\bibitem{becker1987low}
\bibinfo{author}{Becker, H.}, \bibinfo{author}{Rolfs, C.} \&
  \bibinfo{author}{Trautvetter, H.}
\newblock \bibinfo{title}{Low-energy cross sections for 11 b (p, 3$\alpha$)}.
\newblock \emph{\bibinfo{journal}{Z. Phys. A Atomic Nuclei}}
  \textbf{\bibinfo{volume}{327}}, \bibinfo{pages}{341--355}
  (\bibinfo{year}{1987}).

\bibitem{Bonvalet2021Energetic}
\bibinfo{author}{Bonvalet, J.} \emph{et~al.}
\newblock \bibinfo{title}{Energetic $\ensuremath{\alpha}$-particle sources
  produced through proton-boron reactions by high-energy high-intensity laser
  beams}.
\newblock \emph{\bibinfo{journal}{Phys. Rev. E}}
  \textbf{\bibinfo{volume}{103}}, \bibinfo{pages}{053202}
  (\bibinfo{year}{2021}).

\bibitem{Hegelich2023Photon}
\bibinfo{author}{Hegelich, B.~M.}, \bibinfo{author}{Labun, L.},
  \bibinfo{author}{Labun, O.~Z.}, \bibinfo{author}{Mehlhorn, T.~A.} \&
  \bibinfo{author}{Batani, D.}
\newblock \bibinfo{title}{Photon and neutron production as in situ diagnostics
  of proton-boron fusion}.
\newblock \emph{\bibinfo{journal}{Laser Part. Beams}}
  \textbf{\bibinfo{volume}{2023}}, \bibinfo{pages}{e7} (\bibinfo{year}{2023}).

\bibitem{Labaune2013Fusion}
\bibinfo{author}{Labaune, C.} \emph{et~al.}
\newblock \bibinfo{title}{{Fusion reactions initiated by laser-accelerated
  particle beams in a laser-produced plasma}}.
\newblock \emph{\bibinfo{journal}{Nat. Commun.}} \textbf{\bibinfo{volume}{4}},
  \bibinfo{pages}{2041--1723} (\bibinfo{year}{2013}).

\bibitem{Zhang2022Ion}
\bibinfo{author}{Zhang, Y.} \emph{et~al.}
\newblock \bibinfo{title}{{Ion beam stopping power effects on nuclear fusion
  reactions}}.
\newblock \emph{\bibinfo{journal}{Phys. Plasmas}} \textbf{\bibinfo{volume}{29}}
  (\bibinfo{year}{2022}).

\bibitem{nikezic2004formation}
\bibinfo{author}{Nikezic, D.} \& \bibinfo{author}{Yu, K.~N.}
\newblock \emph{\bibinfo{journal}{Mater. Sci. Eng.}}
  \textbf{\bibinfo{volume}{46}}, \bibinfo{pages}{51--123}
  (\bibinfo{year}{2004}).

\bibitem{thomson1911xxvi}
\bibinfo{author}{Thomson, J.~J.}
\newblock \emph{\bibinfo{journal}{The London, Edinburgh, and Dublin
  Philosophical Magazine and Journal of Science}}
  \textbf{\bibinfo{volume}{21}}, \bibinfo{pages}{225--249}
  (\bibinfo{year}{1911}).

\bibitem{carroll2010modified}
\bibinfo{author}{Carroll, D.~C.} \emph{et~al.}
\newblock \emph{\bibinfo{journal}{Nucl. Instrum. Meth. A}}
  \textbf{\bibinfo{volume}{620}}, \bibinfo{pages}{23--27}
  (\bibinfo{year}{2010}).

\bibitem{Nakamura2006Real}
\bibinfo{author}{Nakamura, S.} \emph{et~al.}
\newblock \bibinfo{title}{Real-time optimization of proton production by
  intense short-pulse laser with time-of-flight measurement}.
\newblock \emph{\bibinfo{journal}{Jpn. J. Appl. Phys.}}
  \textbf{\bibinfo{volume}{45}}, \bibinfo{pages}{L913} (\bibinfo{year}{2006}).

\bibitem{Glebov2010National}
\bibinfo{author}{Glebov, V.~Y.} \emph{et~al.}
\newblock \bibinfo{title}{{The National Ignition Facility neutron
  time-of-flight system and its initial performance}}.
\newblock \emph{\bibinfo{journal}{Rev. Sci. Instrum.}}
  \textbf{\bibinfo{volume}{81}}, \bibinfo{pages}{10D325}
  (\bibinfo{year}{2010}).

\bibitem{zhu2017xingguang}
\bibinfo{author}{Zhu, Q.} \emph{et~al.}
\newblock \bibinfo{title}{{The Xingguang-III laser facility: precise
  synchronization with femtosecond, picosecond and nanosecond beams}}.
\newblock \emph{\bibinfo{journal}{Laser Phys. Lett.}}
  \textbf{\bibinfo{volume}{15}}, \bibinfo{pages}{015301}
  (\bibinfo{year}{2017}).

\bibitem{zhang2018angular}
\bibinfo{author}{Zhang, Y.} \emph{et~al.}
\newblock \bibinfo{title}{{An angular-resolved multi-channel Thomson parabola
  spectrometer for laser-driven ion measurement}}.
\newblock \emph{\bibinfo{journal}{Rev. Sci. Instrum.}}
  \textbf{\bibinfo{volume}{89}}, \bibinfo{pages}{093302}
  (\bibinfo{year}{2018}).

\bibitem{vatnitsky1997radiochromic}
\bibinfo{author}{Vatnitsky, S.~M.}
\newblock \emph{\bibinfo{journal}{Appl. Radiat. Isotopes}}
  \textbf{\bibinfo{volume}{48}}, \bibinfo{pages}{643--651}
  (\bibinfo{year}{1997}).

\bibitem{S2003Spectrometry}
\bibinfo{author}{S{\'e}guin, F.~H.} \emph{et~al.}
\newblock \bibinfo{title}{{Spectrometry of charged particles from
  inertial-confinement-fusion plasmas}}.
\newblock \emph{\bibinfo{journal}{Rev. Sci. Instrum.}}
  \textbf{\bibinfo{volume}{74}}, \bibinfo{pages}{975--995}
  (\bibinfo{year}{2003}).

\bibitem{Agostinelli2003G}
\bibinfo{author}{Agostinelli, S.} \emph{et~al.}
\newblock \bibinfo{title}{{GEANT4--simulation toolkit}}.
\newblock \emph{\bibinfo{journal}{Nucl. Instrum. Methods Phys. Res. Section A}}
  \textbf{\bibinfo{volume}{506}}, \bibinfo{pages}{250--303}
  (\bibinfo{year}{2003}).

\bibitem{Fryxell2000FLASH}
\bibinfo{author}{Fryxell, B.} \emph{et~al.}
\newblock \bibinfo{title}{{FLASH: An adaptive mesh hydrodynamics code for
  modeling astrophysical thermonuclear flashes}}.
\newblock \emph{\bibinfo{journal}{Astrophys. J. Suppl. Ser.}}
  \textbf{\bibinfo{volume}{131}}, \bibinfo{pages}{273--334}
  (\bibinfo{year}{2000}).

\bibitem{Salpeter1954ElectronSA}
\bibinfo{author}{Salpeter, E.}
\newblock \bibinfo{title}{Electron screening and thermonuclear reactions}.
\newblock \emph{\bibinfo{journal}{Aust. J. Phys.}}
  \textbf{\bibinfo{volume}{7}}, \bibinfo{pages}{373--388}
  (\bibinfo{year}{1954}).

\bibitem{Lv2022Enhanced}
\bibinfo{author}{Lv, W.}, \bibinfo{author}{Duan, H.} \& \bibinfo{author}{Liu,
  J.}
\newblock \bibinfo{title}{Enhanced proton-boron nuclear fusion cross sections
  in intense high-frequency laser fields}.
\newblock \emph{\bibinfo{journal}{Nucl. Phys. A}}
  \textbf{\bibinfo{volume}{1025}}, \bibinfo{pages}{122490}
  (\bibinfo{year}{2022}).

\bibitem{Estabrook1978Qualitative}
\bibinfo{author}{Estabrook, K.}
\newblock \bibinfo{title}{Qualitative aspects of underdense magnetic fields in
  laser-fusion plasmas}.
\newblock \emph{\bibinfo{journal}{Phys. Rev. Lett.}}
  \textbf{\bibinfo{volume}{41}}, \bibinfo{pages}{1808--1811}
  (\bibinfo{year}{1978}).

\bibitem{Willi1981Optical}
\bibinfo{author}{Willi, O.}, \bibinfo{author}{Rumsby, P.} \&
  \bibinfo{author}{Sartang, S.}
\newblock \bibinfo{title}{Optical probe observations of nonuniformities in
  laser-produced plasmas}.
\newblock \emph{\bibinfo{journal}{IEEE J. Quantum Electron}}
  \textbf{\bibinfo{volume}{17}}, \bibinfo{pages}{1909--1917}
  (\bibinfo{year}{1981}).

\bibitem{Gao2012Magnetic}
\bibinfo{author}{Gao, L.} \emph{et~al.}
\newblock \bibinfo{title}{Magnetic field generation by the rayleigh-taylor
  instability in laser-driven planar plastic targets}.
\newblock \emph{\bibinfo{journal}{Phys. Rev. Lett.}}
  \textbf{\bibinfo{volume}{109}}, \bibinfo{pages}{115001}
  (\bibinfo{year}{2012}).

\bibitem{SID4-HR}
\bibinfo{title}{{See http://phasicscorp.com/cameras/sid4-hr/ for information of
  the {SID4-HR} sensor}} .

\bibitem{tendl}
\bibinfo{title}{{See https://tendl.web.psi.ch/tendl2021/tendl2021.html}} .

\end{thebibliography}

\end{document}